\documentclass[12pt,preprint]{aastex}

\newcommand{\autoref}[1]{{\it #1}}
\newcommand\kms{\,km\thinspace s$^{-1}$ }  

\shorttitle{SAPAC}
\shortauthors{Dunn \& Jerjen}

\begin{document}

\title{First results from SAPAC: towards a 3D-picture of the Fornax cluster core\footnote{Based on observations collected at the ESO-VLT, under program ESO 68.A-0176}}

\author{Laura P. Dunn \and Helmut Jerjen}

\affil{Research School of Astronomy and Astrophysics, 
Mt Stromlo Observatory, Australian National University, 
Cotter Road, Weston ACT 2611 Australia }

\email{laurad@mso.anu.edu.au, jerjen@mso.anu.edu.au}

\begin{abstract}
A sophisticated SBF analysis package has been developed, designed to measure distances of early-type galaxies by means of surface brightness fluctuations of unresolved stars. This suite of programs called SAPAC is made readily available to the astronomical community for extensive testing with the long-term goal to provide the necessary tools for systematic distance surveys of early-type galaxies using modern optical/NIR telescopes equipped with wide-field cameras. We discuss the technical and scientific concepts of SAPAC and demonstrate its capabilities by analysing deep $B$ and $R$-band CCD images of 10 dwarf elliptical (dE) galaxy candidates in the Fornax cluster obtained with FORS1 at the VLT. All candidates are confirmed as cluster members. We then turn our attention to the innermost region of the Fornax cluster. A total of 29 early-type galaxies closer than three cluster core radii ($2^\circ$) from the central galaxy NGC1399 have radial velocities and SBF distances.  Their Hubble diagram exhibits a pronounced S-shaped infall pattern suggesting a picture that Fornax is still in the process of formation during the present epoch through a general collapse and possible accretion of distinct groups of galaxies. From fitting a model we estimate the cluster mass within 720\,kpc projected distance from NGC1399 which is $2.3\pm0.3\times 10^{14}\mathcal{M}_\odot$. The associated collapse time is $t_{\rm collapse}=2.9{+1.6 \atop -0.9}$\,Gyr. 
After cleansing our galaxy sample from a few kinematical outliers the true distance of the Fornax cluster core is 
determined at 20.13$\pm$0.40\,Mpc [$(m-M)_0=31.51\pm 0.04$\,mag]. Applying a bootstrap resampling 
technique on the distance distribution with individual distance errors taken into account further reveals a small 
intrinsic cluster depth of $\sigma_{\rm int}=0.74{+0.52 \atop -0.74}$\,Mpc in best agreement with the cluster's 
linear extension in the sky: $\sigma_{\rm R.A}=\sigma_{\rm DEC}\approx0.5$\,Mpc. We conclude that the early-type 
galaxy population in the Fornax cluster must be spatially well constrained with no evidence of elongation along 
the line-of-sight in contrast to the Virgo cluster. Moreover, we find marginal evidence for substructure, a results 
that is consistent with the young evolutionary state of the cluster and the overall galaxy infall. Combining the kinematically defined 
cluster distance with the mean cosmological velocity for the central cluster galaxy sample yields a 
Hubble constant of $H_0 = 63\pm 5$ \kms Mpc$^{-1}$. 
\end{abstract}


\maketitle

\section{Introduction}
Finding low star density dwarf galaxies in the local Universe is, by nature, a challenging task. Over the last two decades, numerous surveys \citep[e.g.][]{B85,J97,F90,JBF00,K95,CA98} have produced hundreds of candidates in nearby groups and clusters based on morphological properties. But due to the lack of any distance information large numbers still remain only dwarf {\em candidates} today. 

To separate the wheat from the chaff, i.e., genuine dwarf galaxies from similar looking background spirals and giant ellipticals, distances are required. The importance of distances, however, extends far beyond the unambiguous identification and spatial location of new dwarf galaxies. The careful construction of highly complete, volume-limited galaxy samples like the 10\,Mpc sphere \citep{K79,S92,K04} yields the exact number and distribution of dwarfs relative to larger galaxies and provides crucial empirical input to better constrain theories of galaxy formation, including the faint-end of the galaxy luminosity function and the merging rate \citep{TS02,JTT04}. Furthermore, accurate distances permit studies of the dwarf galaxy distribution in 3-D and, if combined with kinematical and stellar population data, provide insight into the importance of environmentally driven over self-induced physical processes that govern dwarf galaxy evolution \citep[e.g.][]{PP01}.

Unlike the scarcely evolved, low density environment populating dwarf irregular galaxies, dwarf elliptical (dE) galaxies are lacking significant amounts of neutral hydrogen gas preventing a relatively simple distance estimate via 21cm\,redshift. Furthermore, little help is coming from empirical scaling relations between galaxy luminosity and surface brightness \citep{BC91} or S\'ersic shape parameter \citep{B98} as the large intrinsic scatter turns out to be unsuitable for individual distance measurements.  That is why in the last decade a formidable observational effort has been devoted to deriving distances to dEs by using the tip magnitude of the red giant branch (TRGB), a powerful Pop II distance indicator \citep[][and references therein]{K03,D90}. However, the requirement of accurate photometry for individual stars 1--2 magnitudes below the tip [$m_I\geq 25.8+5\log(D/5 {\rm Mpc})$] at distance $D$[Mpc] sets a practical application limit at $\approx$ 5\,Mpc beyond which TRGB measurements quickly become very costly and time consuming. This was demonstrated by \cite{H98} who spent nine hours with the HST to obtain the TRGB distance for a single dE galaxy (IC3388) in  the Virgo cluster. 

Because of this limitation, a new distance indicator, the so-called surface brightness fluctuation (SBF) method, has received increasing attention in recent years. The method quantifies the statistical pixel-to-pixel variation of star counts across a galaxy image with the major technical advantage of working on unresolved stellar populations. Ground-based CCD images with relatively short exposure times can be used instead of high resolution space-based observations making the SBF method extremely efficient. Distance measurements for early-type galaxies far beyond the reach of any other distance indicator become possible. 

The theoretical framework of the SBF method was developed by \cite{T88} to study the distribution of luminous giant ellipticals in the local Universe \citep[e.g.][]{T01}. While SBF applications were initially focused on high surface brightness systems \citep{A94,J96,T97}, Jerjen and collaborators \citep{J98, J00, J01,J03} demonstrated that the method works equally well for dEs with stellar surface densities as low as $\mu_{\rm B_{eff}} =26$\,mag\,arcsec$^{-2}$ and luminosities comparable to Local Group dwarfs like Leo\,I, And\,II, and Sculptor ($M_B= -10$\,mag). The SBF method has since been successfully applied to dwarf candidates in the distance range from 10--40\,Mpc, employing 2-m ground-based telescopes \citep{R05}, the 8-m ESO-VLT \citep{M03,J04}, and the HST \citep{Jo05, M05}.

Noticing that dE galaxies are by far the most numerous galaxy type at the current cosmological epoch \citep{F94}, the SBF technique, in combination with modern wide-field imaging at optical and near-IR telescopes like SUBARU, VISTA, and LSST, has the potential to play a major role in future {\em distance} surveys of early-type galaxies in the local Universe. Using extended source catalogues from SDSS and Pan-STARRS in the North and SkyMapper in the South there is the prospect of carrying out SBF distance surveys in a similar fashion as redshift surveys are done today. For the efficient processing of such large amount of imaging data we are currently engaged in a project of developing an automated SBF Analysis Pipeline for the Astronomical Community called SAPAC. 

In \S\ref{sapac} we will introduce SAPAC, discuss the data requirements for the pipeline and give an overview of the SBF method.  We will then demonstrate its capabilities in \S\ref{test} by processing VLT+FORS1 CCD images for 10 Fornax cluster dE candidates of similar quality as existing and future survey telescopes are expected to produce. In \S\ref{discuss} we endeavor to extract clues about the 3-D cluster structure from the spatial distribution of early-type galaxies in the central region of Fornax. Combining galaxy velocities and distances and analysing the Hubble diagram enables the measurement of various cluster parameters such as the total mass and dynamical collapse time, as well as a value for the Hubble constant. We summarize our results in \S\ref{conc}.

\section{SAPAC description} \label{sapac}
Previous SBF measurements have entailed individuals hand selecting regions in galaxy images for the analysis.  To make the data reduction more efficient and results as impartial as possible  we have developed a rapid and easy-to-use visual SBF tool. SAPAC\footnote{\url{http://www.mso.anu.edu.au/$\sim$laurad/sapac}} is a software package that addresses the entire SBF analytical process from galaxy modeling and Fourier analysis to fluctuation magnitude calibration  and presentation of the distance results. It works on any pair of CCD images that meets four basic criteria as discussed in the following paragraph. SAPAC currently runs under Linux but will be made available for other operating systems in the near future. It comprises of a suite of Perl scripts using an IRAF module Astro::IRAF::CL\footnote{\url{http://search.cpan.org/$\sim$sjquinney/Astro-IRAF-CL-0.2.0/CL.pm}} and a sophisticated graphical user interface, also written in Perl, employing Tk\footnote{http://search.cpan.org/$\sim$ni-s/Tk/}. For more information on the computing requirements to run the software we refer to the SAPAC manual.

\subsection{Data Requirements} \label{require}
There are four checks that help the SAPAC user to decide whether a set of CCD images of a galaxy are suitable for the SBF analysis. They are as follows: 
\smallskip

\noindent{\em Galaxy morphology:}
the galaxy light distribution must be overall radially symmetric and have minimal structure to permit accurate 2-D modeling. This technical constraint basically restricts the application to early-type galaxies and bulges of spiral galaxies. We like to note that sophisticated galaxy fitting programs such as {\em galfit} \citep{P02} can actually successfully model entire spiral galaxies. But even so, the SBF signal depends on the star formation history, age, and metallicity of the underlying stellar population and thus is less constrained for spiral than for elliptical galaxies. Working with a more complex parameter space leads to a larger scatter in the calibration of the SBF method. Hence, there is a stellar population argument too why an SBF analysis should be restricted to early-type galaxies, but this might change in the future.
\smallskip

\noindent{\em Photometric bands:} As the measured fluctuation magnitude is colour dependent calibrated CCD images in two different photometric bands are required, e.g.~($B$, $R$) or ($g_{475}$, $z_{850}$).
\smallskip

\noindent{\em Seeing:} In essence, the CCD images must be obtained under good atmospheric conditions. A seeing value of 1\,arcsec or better is recommended for the image used for the SBF analysis (i.e. $R$ or $z_{850}$), but this depends on the angular size of the galaxy. A rule of thumb is  $FWHM\le r_{eff}[\arcsec]/20$, where $r_{eff}$ is the half-light radius of the galaxy.  The image in the other photometric band can be of slightly lower quality but should not exceed 20-30 percent to remain immune to the effect of pixel smearing.
\smallskip

\noindent{\em Integration time:} the total integration time $t$ for the image in the band used for the SBF analysis is estimated from five quantities: 
  \begin{itemize}
  \item{mean surface brightness of the galaxy ($\mu_{gal}$)}
  \item{surface brightness of the sky background ($\mu_{sky}$)}
  \item{distance modulus of the galaxy ($DM$)}
  \item{fluctuation luminosity of the underlying stellar population in the SBF band ($\overline{M}$)}
  \item{photometric zero point of the telescope+instrument system, i.e.~the magnitude of a star providing 1 ADU\,s$^{-1}$ on the CCD detector ($m_1$)}
  \end{itemize}
The integration time can then be expressed as a function of the  signal-to-noise ratio S/N in the derived SBF power spectrum (see \S\ref{sapacdescrip}):
\begin{equation}
t=S/N\cdot 10^{0.4(\mu_{gal}-\mu_{sky}+DM+\overline{M}-m_1)}
\label{int}
\end{equation}

\begin{figure}
\begin{center}
\includegraphics[totalheight=0.6\textheight]{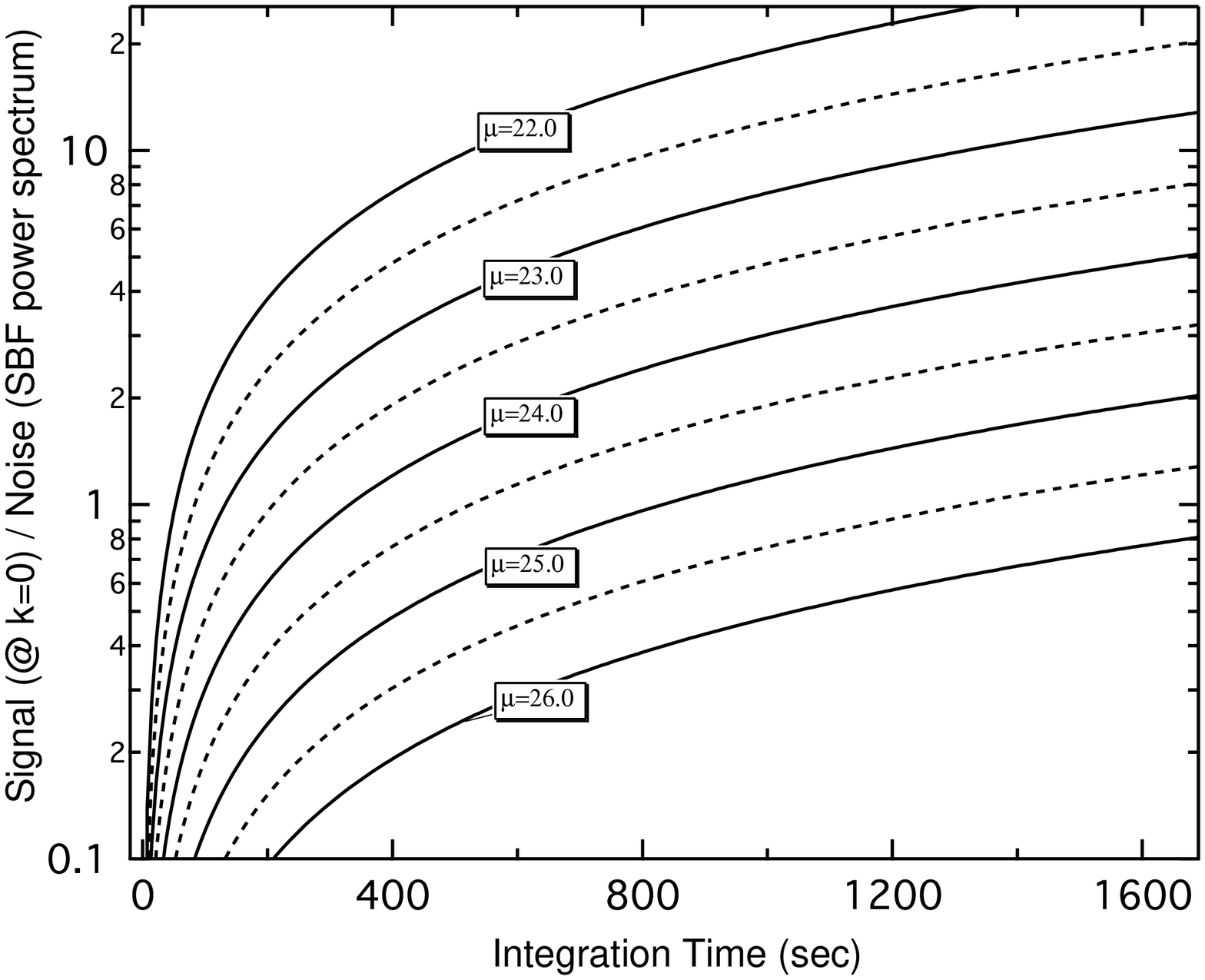}
\caption{An illustration showing how the signal-to-noise in the SBF power spectrum increases with length of exposure time and galaxy surface brightness at the adopted distance modulus of 31.50\,mag. The modeled telescope is the VLT+FORS1 with a photometric zero point of $m_{1,R}=$27.10\,mag and a sky brightness of $\mu_{sky,R}=20.8$\,mag\,arcsec$^{-2}$. The surface brightness fluctuation luminosity is set at $\overline{M}_R=-1.30$\,mag.\label{inttime}}
\end{center}
\end{figure}

 Fig.\ref{inttime} illustrates Eq.\ref{int}, plotting signal-to-noise versus integration  time for different values of $\mu_{gal}$. This example is tuned for an old, metal-poor  stellar system ($\overline{M}_R=-1.30$\,mag)  at the tentative distance modulus  of 31.50\,mag,  observed in the $R$-band with  the VLT at no  moon condition ($m_{1,R}=$27.10\,mag, $\mu_{sky,R}=20.8$\,mag\,arcsec$^{-2}$).  From experience, a SBF distance can be derived reliably when the offset between  shot noise level and SBF signal at wave number $k=0$ is 0.5 dex or  higher \citep[see][Fig.~8] {R05}. To achieve the equivalent S/N=3 in the  power spectrum,  the minimum integration time for a Fornax cluster dE with  $\mu_{gal}$=24.5 mag arcsec$^{-2}$ is 1600 seconds.  This is a factor 20 shorter that the 32,000 seconds spent by \citet{H98} to measure the TRGB distance for the Virgo dwarf IC3388 with HST.

\subsection{SAPAC outline}\label{sapacdescrip}
The SBF reduction pipeline follows the analysis steps as described by \citet{J00,J01,J04}, i.e., galaxy modeling, SBF field selection, Fourier analysis, fluctuation magnitude calibration, estimate of fluctuation contributions from unwanted sources, distance measurement. All steps are combined into a single package with a graphical user interface that is designed to be easy-to-use for novices but is powerful enough for experienced users. The pipeline requires an astronomical imaging and data visualization application like DS9 \citep{JM03} to be running and that the CCD data available meets the requirements outlined in the previous section.  Although it is necessary to have observations in two bands to measure a distance, SAPAC can calculate the power spectra of selected galaxy fields on a single CCD frame.  This is useful to test the quality of a image for the purpose of an SBF analysis and thus to decide whether applying for more observing time to get the missing second passband image  would be productive.  

Obviously, the CCD images must be photometrically calibrated and 
carefully sky-subtracted, and the photometric zero points available as keywords in the fits headers. 
Furthermore, the two images must be on the same pixel scale and registered. Fig.~\ref{steps} shows the steps by which the process follows for one of our observed galaxies FCC318.

\subsubsection{Modeling the Light Distribution}\label{model}
\begin{figure}
\begin{center}
\includegraphics[totalheight=0.7\textheight]{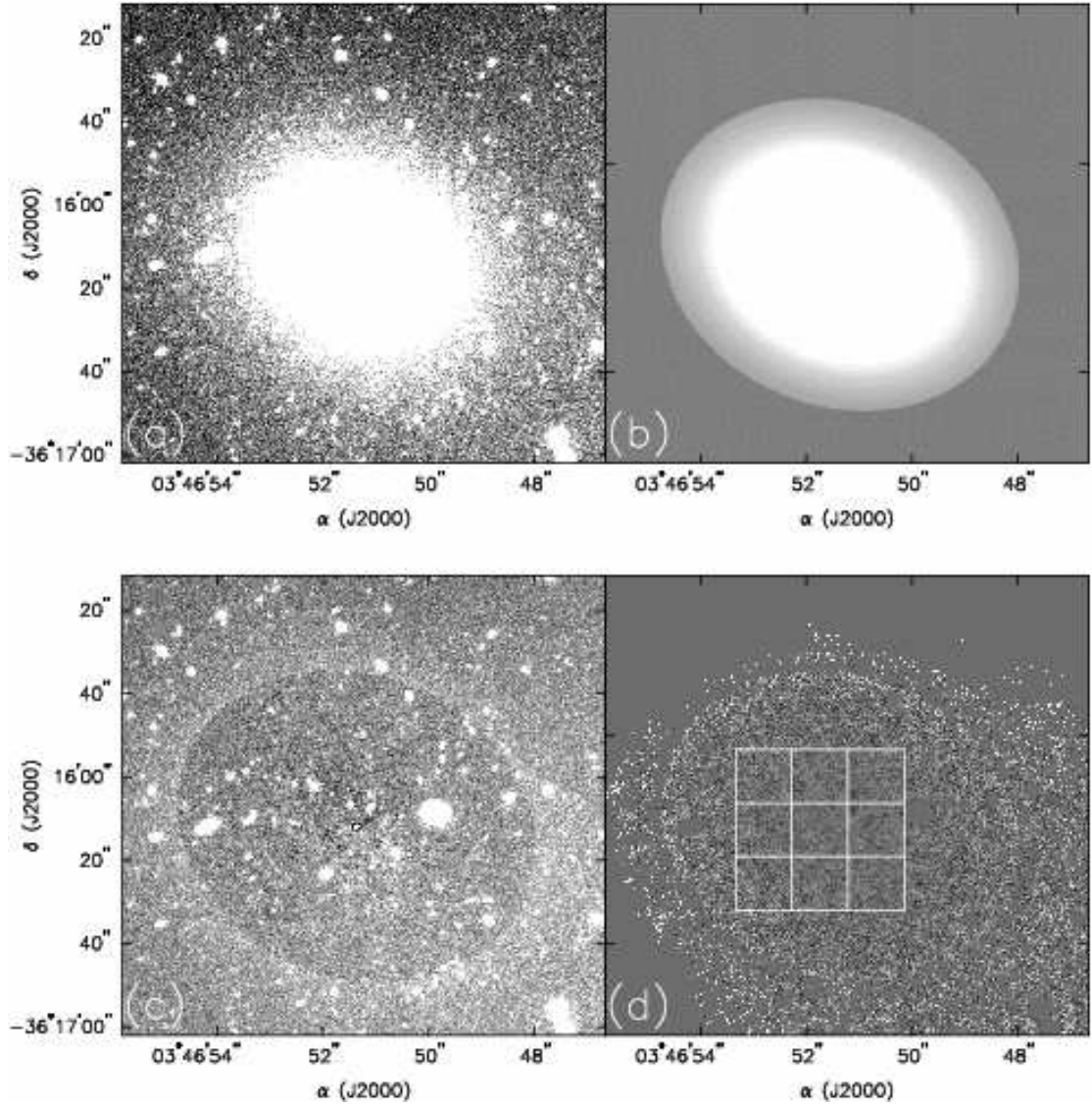}
\caption{The four major steps leading up to the individual field analysis in SAPAC. This example shows FCC318. (top left, a) shows the original FORS1 $R$-band image of the galaxy; (top right, b) the best-fitting model of the light distribution from SAPAC; (bottom left, c) the residual, after subtraction of the model; (bottom right, d) the final result after normalization, masking and a $3\times3$ grid is overlaid. The mottling of the unresolved stars in the galaxy center is clearly visible.\label{steps}}
\end{center}
\end{figure}

The initial step in the SBF analysis process is to accurately model the light distribution of the galaxy (Fig.~\ref{steps}a) in both passbands.  SAPAC finds the luminosity-weighted galaxy center from a user-defined square box and employing the IRAF command  $imcentroid$. The IRAF procedure $ellipse$ is used to generate the model which is subsequently subtracted from the original image. The best-fitting model and residual image for FCC318  is shown in Figs.~\ref{steps}b and \ref{steps}c.  The same modeling constraints are used on the second image.  The 2-D modeling of the light distribution is a crucial step in preparation for the SBF analysis as any residual may affect the final results. Consequently, this delicate part is currently the most interactive in the process. Although the original purpose of this part of the software is to fit galaxies, the algorithm can be also used in a broader context of morphology studies or for deblending objects.

\subsubsection{Normalisation, Masking and PSF Analysis}\label{norm}
In the next stage, the residual image of the SBF band from step one is divided by the square root of the model for flux normalisation and bright objects are masked out.  The masking is done automatically by SAPAC at two separate cut-off levels. The first mask is applied to high pixel values with a large radius surrounding these peaks in order to remove bright foreground stars; the second mask is applied at a lower cut with a smaller radius to excise fainter foreground stars and background galaxies. If by chance a part of a galaxy is heavily masked it should be omitted for the SBF analysis. An example of the normalized, masked image is seen in Fig.~\ref{steps}d, with a grid overlaid which will be discussed in the next section.

The third step is concerning the reference star that is needed to determine the PSF convolved component in the power spectra of the SBF fields.  A suitable star selected by the user should be bright and well isolated, such that light from the target galaxy is not affecting it.  We also recommend that the star is relatively close to the galaxy so that the influence of possible spatial PSF variation across the CCD chip is kept to a minimum.  The profile of the star is determined by fitting a point spread function model and this will be used for the decomposition of the galaxy power spectrum.

\subsubsection{SBF field analysis}\label{anal}
The fourth step gives the user the opportunity to define the number of square SBF fields and their approximate size. An area of the galaxy is selected over which the grid is placed, as seen in Fig.\ref{steps}d. Each field is then analysed individually. The analysis comprises of each of the fields being Fourier transformed and the azimuthally averaged power spectrum being computed. The power spectrum of the mask is subtracted to remove the effects of the masking from the analysis.  SAPAC then fits a linear combination of the flux normalized and exposure time weighted PSF power spectrum of the star and a constant at the power spectrum of the SBF field ${\rm PS}(k) = P_0\cdot {\rm PS}_{\rm star}(k)+P_1$, demanding a least squares minimisation. Data points at low spatial frequencies ($k \leq 5$) are omitted as they are likely affected by imperfect galaxy model subtraction. The derived power, $P_0$, is still mildly contaminated by a contribution of unresolved background galaxies $P_{\rm BG}$. To estimate this contribution SAPAC employs the \citet{Je98} equation adjusted for the $R$-band \citep{J01}. Another potential source of unwanted fluctuations are globular cluster systems. Although this is an important issue for giant ellipticals, contributions from globular clusters in dEs are negligible due to the very low GC numbers observed in dEs \citep{ML98}.  The expected few would be excised during the masking process and hence no further correction on $P_0$  is applied.  

When corrected for background galaxies and Galactic extinction, $A_R$, the amplitude $P_0$ relates to the fluctuation magnitude of the unresolved stars in the galaxy ($R$-band) as: $\overline{m}_{\rm stars}=m_{1}-2.5\log_{10}((P_0-P_{\rm BG})/t)-A_R$, where $t$ is the exposure time and $m_1$ is the photometric zero-point. For the purpose of calibrating the derived fluctuation magnitude, the colour $(B-R)_0$ of each SBF field is finally calculated by SAPAC using the galaxy model in the second passband.

\subsubsection{Calibration and distance measurement} \label{calib}
A semi-empirical calibration of the fluctuation magnitude $\overline{M}_R$ for the colour range $0.8<(B-R)_0<1.5$ was established \citep{J05} by combining  observational results from 36 SBF fields in seven nearby dwarf galaxies \citep{J01, R05} with independent distance measurements available from TRGB \citep{K99,K00} or SBF \citep{T01} and extending it with theoretical results from stellar population synthesis models by \citet{W94} and evolutionary tracks from the Padova library \citep{B94}. Two distinct relationships were found depending on composition, age, and metallicity of the underlying stellar population: a parabolic branch covering the colour  range $0.8 < (B-R)_0 < 1.35$ and a linear branch covering  $1.1 < (B-R)_0 < 1.5$. The two calibration loci are described  analytically by Eqs.~(2)+(3) with an intrinsic scatter of 0.15\,mag. Both are used by SAPAC for the  distance measurement:
\begin{eqnarray}
\label{calibeq}
\overline{M}_R&=& 6.09(B-R)_0-8.94\\
\overline{M}_R&=& 1.89[(B-R)_0-0.77]^2-1.39
\end{eqnarray}
Future versions of SAPAC will have the SBF calibration for the Sloan filter $z_{850}$ implemented as presented by \citet{M05}.

SAPAC analyses each field automatically but it will be up to the user to keep or discard results from individual fields for the distance determination. Two of the most common reasons for omitting fields are imperfection in the modeling, especially around the center of the galaxy or fields with a high percentage of masked pixels due to superimposed bright foreground stars. Once the user has decided on the fields to include in the distance measurement, the pipeline will output the power spectrum of each field, a plot of the selected fields over the galaxy, and will generate the best-fitting calibration diagram with the distance modulus and estimates error including uncertainties in photometric zero points, power spectrum fitting, sky level determination, and Galactic extinction. Examples of the output will be shown in the next section.

\section{Testing the pipeline on Fornax cluster dE candidates} \label{test}
As an opportunity to test the performance of SAPAC we processed CCD images of ten dE candidates from the Fornax Cluster Catalog \citep[FCC, ][]{F89}. The basic properties of these galaxies are listed in Table \ref{basicdata}, where Column (1) is the FCC number, (2) is the morphological type, (3) and (4) are RA and Declination for epoch J2000, (5) is the integrated $B$-magnitude, (6) is the heliocentric velocity, (7) and (8) are the Galactic extinction in the $B$ and $R$-bands, references are listed below the table.  The velocities available in the literature for eight of the ten galaxies suggest cluster membership but no distances were measured to date. FCC279 and FCC318 even lack a velocity.

\subsection{Observations and Reduction}
The FORS1 (FOcal Reducer/Low dispersion Spectrograph) at UT3 of the Very Large Telescope at ESO Paranal Observatory was used in service mode to obtain deep $B$ and $R$-band images of the target galaxies. The CCD was read out by four amplifiers, each reading one quarter of the CCD, known as four-port mode. The galaxies were centered on one of these four quadrants, thus leaving the remaining three empty and available to be used as night flats.  Each galaxy, except FCC274, was observed in 3 sets of 500\,s exposures in the $B$-band and 700\,s exposures in the $R$-band, each with slightly different pointings. As the observing log in Table~\ref{logbook} shows, FCC274 only has one exposure in the $R$-band but since the seeing in that single frame was excellent ($0.56\arcsec$) we were able to continue with our analysis on this galaxy.  Galaxies FCC245 and FCC252 were observed three times, totaling 9 observations, but this was due to poor photometric conditions on some nights.  We chose to use only the three images from the night with the best photometric conditions.

The data was reduced using routines within IRAF\footnote{IRAF is distributed by the National Optical Astronomy Observatories, which is operated by the Association of Universities for Research in Astronomy, Inc., under contract with the National Science Foundation}. The bias level was removed from each of the quadrants using the overscan regions and a masterbias that was created from bias frames taken before and after every observing night.  Night flats were median-combined to create masterflats for each filter and subsequently the images were flatfield corrected.

A series of photometric standard stars \citep{L92} were observed in various passbands under different airmasses every night in service mode.  Due to these frequent observations, trends for the colour terms and extinction coefficients at Paranal are accurately monitored and reported by the VLT Quality Control and Trending Services \citep{HS02} at Garching.  Table~\ref{cterm} shows the extinction coefficients for $B$ (Column 2) and $R$ (Column 3) bands and the colour term (Column 4) derived for each night (Column 1).  These values and our own set of standard stars were used to determine the photometric zero points in Column 5 (see Table \ref{cterm}).

Using the $starfind$, $xyxymatch$ and $imalign$ commands all the images from a particular galaxy were aligned by matching the positions between 30 to 50 stars.  Star-free areas distributed uniformly over the CCD image but away from the galaxy were selected for an accurate sky background determination.  The sky subtracted images from the same passband were then median combined using $imcombine$ and finally these master images were flux calibrated.

\subsection{Distances derived using SAPAC}\label{results}
\begin{figure}
\begin{center}
\includegraphics[width=6.cm,angle=270]{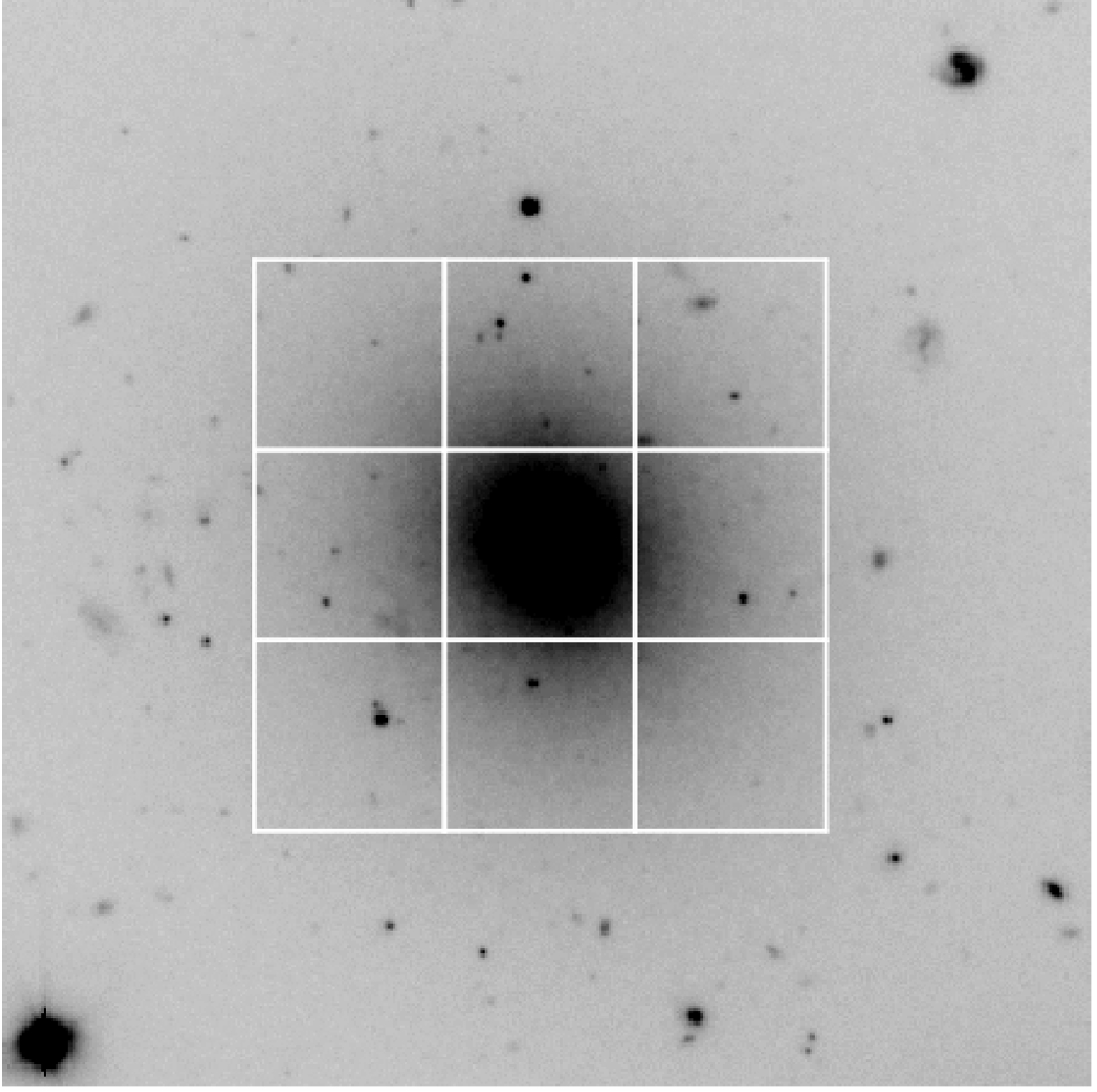}\hfill
\includegraphics[width=6.cm,angle=270]{f3b.eps}\\
\vspace{0.5in}
\includegraphics[width=6.cm,angle=270]{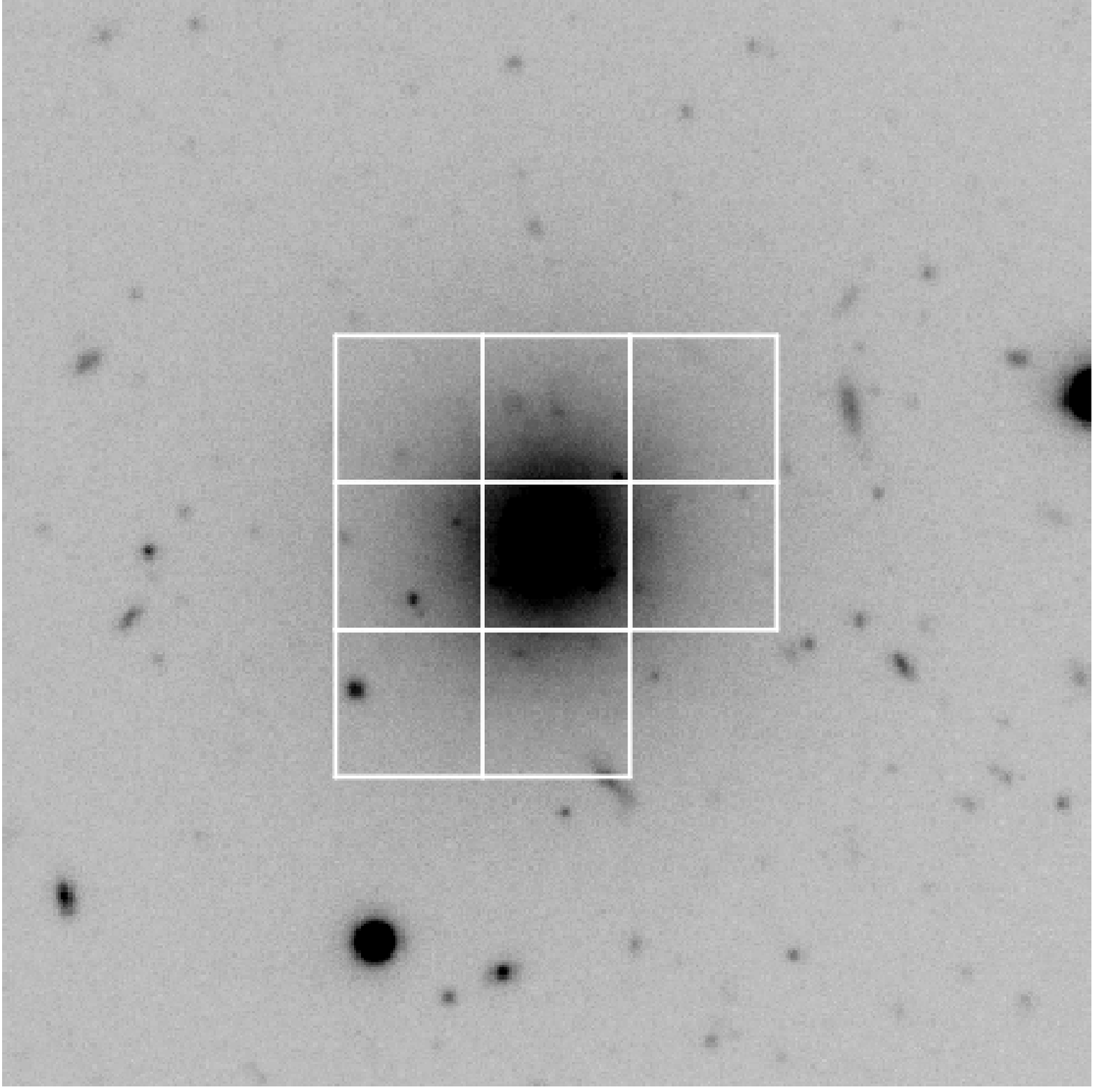}\hfill
\includegraphics[width=6.cm,angle=270]{f3d.eps}
\caption{$R$-band images for the galaxies FCC222 (top) and FCC243 (bottom) with the grid of user-selected SBF fields superimposed. The corresponding power spectra are shown on the right. The observational data (triangles) are best fitted by the sum of a scaled version of the PSF power spectrum and a constant (dashed line).
\label{galgrids}}
\end{center}
\end{figure}

The SBF analysis of the 10 Fornax dE candidates listed in Table~\ref{basicdata} was performed using SAPAC. The chosen SBF fields for two of the sample galaxies are shown in Fig.~\ref{galgrids} for illustration purpose.  As mentioned before, fields that were not included in the final analysis were either the center fields, which had residual from the modeling and subtraction, or, as in the case of FCC207, a nearby star created difficulties in determining the center, or they were outer fields which had large amounts of masked areas or the fluctuation signals were too low, meaning poor fitting of the power spectrum.

Table~\ref{fieldparam}, serving as an example, summarises the measured values in the SBF analysis for the two galaxies shown in Fig.~\ref{galgrids}, i.e. FCC222 and FCC243: galaxy name and SBF field ID (Column 1), the pixel size of the SBF fields (Column 2), $m_1$, the magnitude of a star providing 1 ADU\,s$^{-1}$ on the CCD (Column 3), the mean galaxy surface brightness in ADU (Column 4), the sky surface brightness in ADU (Column 5), $P_0$, the time normalized amplitude of the best fit at wavenumber k=0 with the error in brackets (Column 6), $P_1$, the level of the scale free white noise component in the power spectrum (Column 7),  the signal-to-noise where S/N =$(P_0-P_{BG})/(P_1+P_{BG})$ (Column 8), and the fractional amount of fluctuation from unresolved high redshift galaxies, $P_{BG}/P_0$ (Column 9).  The S/N for all ten galaxies ranges from 2.9 to 15.0 with a median value of 6.8. Most of the fields have a contribution from the background galaxies between 3 and 5\%, with the exception of FCC252 which has a significantly higher fraction, at 8-10\%.  The fluctuation magnitude $\overline{m}_R^0$, corrected for the background galaxy component and foreground extinction \citep{S98}, is listed for each field in column 3 of Table~\ref{fieldmag}, again for the two selected galaxies.  Column 4 of this table gives the measured $(B-R)_0$ colour for each field of each galaxy, also corrected for foreground extinction.  The power spectrum for each selected field for FCC222 and FCC243 are shown in Fig.~\ref{galgrids}, where the layout of the power spectrum corresponds to the fields placed on the galaxy.  We note that the level of the scale free white noise component in the power spectra, equivalent to the ratio of sky and galaxy surface brightness, varies as expected. It is the lowest in the galaxy center, where the surface brightness peaks, and is significantly higher in the outer fields, where the star density is low.

\begin{figure}
\begin{center}
\includegraphics[height=0.75\textheight]{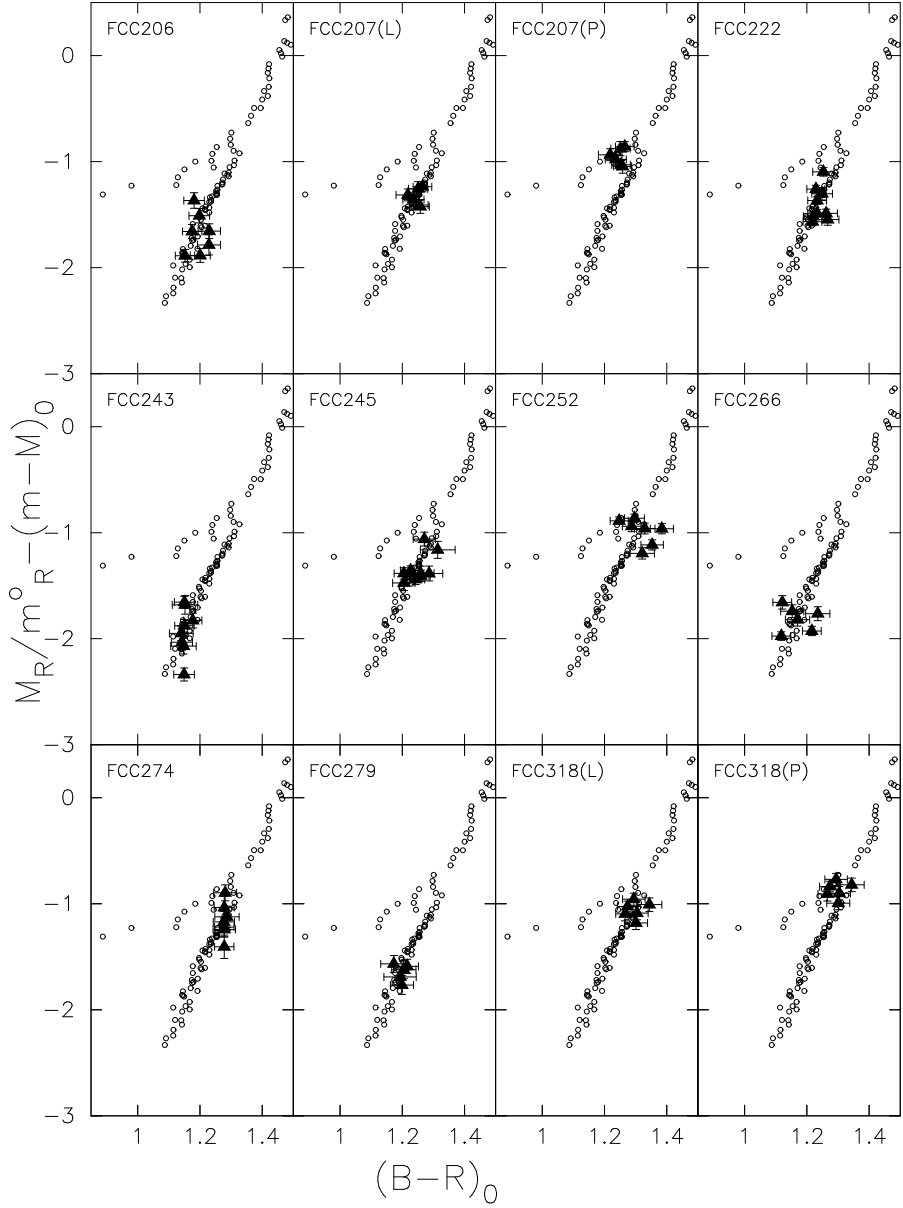}
\caption{The calibrated sets of fluctuation magnitudes (filled triangles) with error bars for the ten Fornax dEs analysed.  The open circles are the 170 data points that define the semi-empirical calibration \citep[see][]{J05}.  In two cases (FCC207 and FCC318) results for both branches (labeled ``L" for linear and ``P" for parabolic) are shown as we could not confidently discard one of the two possible results. For the distance values see Table~\ref{distlist}. \label{calibfigs}}
\end{center}
\end{figure}

Table~\ref{distlist} lists the SBF distances for the 10 galaxies analysed, with two results for FCC207 and FCC318.  The second result for both these galaxies arises from using the parabolic calibration branch as, due to the $(B-R)_0$ colours of the fields being close to the intercept of the two branches, both calibration branches provided appropriate fits.  These results are graphically shown in Fig.~\ref{calibfigs}, where 12 calibration graphs can be seen, referring to the 10 individual galaxies plus the 2 parabolic branch options, denoted by ``P".  
As a first result, we find that the observed distance range of $17.5$\,Mpc $<D<23.8$\,Mpc confirms all sample Fornax dE candidates as cluster members including FCC279 and FCC 318 which do not have velocity measurements to date. 

The error in the power spectrum fitting is between 2 and 11\% for all galaxies. The remaining sources of errors are 2\% for the PSF normalization, 1-2\% for the variation in shape of the stellar PSF over the CCD, 0.02\,mag in $R$ and 0.03\,mag in $B$ for the photometric calibration uncertainty and 16\% error in foreground extinction \citep{S98}.  Hence the combined error for an individual $\overline{m}_R^0$ measurement is between 0.03\,mag and 0.11\,mag (Column 3 in Table~\ref{fieldmag}). Column 4 shows the error in the local colour of each field and this was determined using errors in sky level measurements, photometric zero points and Galactic extinction.

\section{Discussion}\label{discuss}
\subsection{Distance distribution and Substructures}\label{distsub}
With the availability of significant numbers of distances for cluster galaxies true 3-D substructure studies in clusters have finally come of age. Most recent distance studies of the Virgo cluster have found extensive substructure \citep{W00,S02,J03}. \cite{W00} used SBF distance measurements from \cite{T01} to 14 elliptical galaxies in the northern part of the Virgo cluster to show a strong collinear arrangement in three dimensions, defining a principal axis between NGC4387 and NGC4660, otherwise known as cluster A \citep{B87}.  Cluster B is centered around M49 and these two major concentrations of the Virgo Cluster are thought to be surrounded by three galaxy clouds. \cite{J04} also used the SBF method to determine distances to 16 dE galaxies in the Virgo cluster and found first evidence for line-of-sight bimodality in cluster A, when combined with 24 ellipticals with SBF distances from \cite{T01} and \cite{N00}.  They concluded that the northern cluster A consists in fact of two dynamically distinct systems, a small group around M86 falling into the M87 subcluster from the back. 

\begin{figure}
\begin{center}
\includegraphics[totalheight=0.6\textheight]{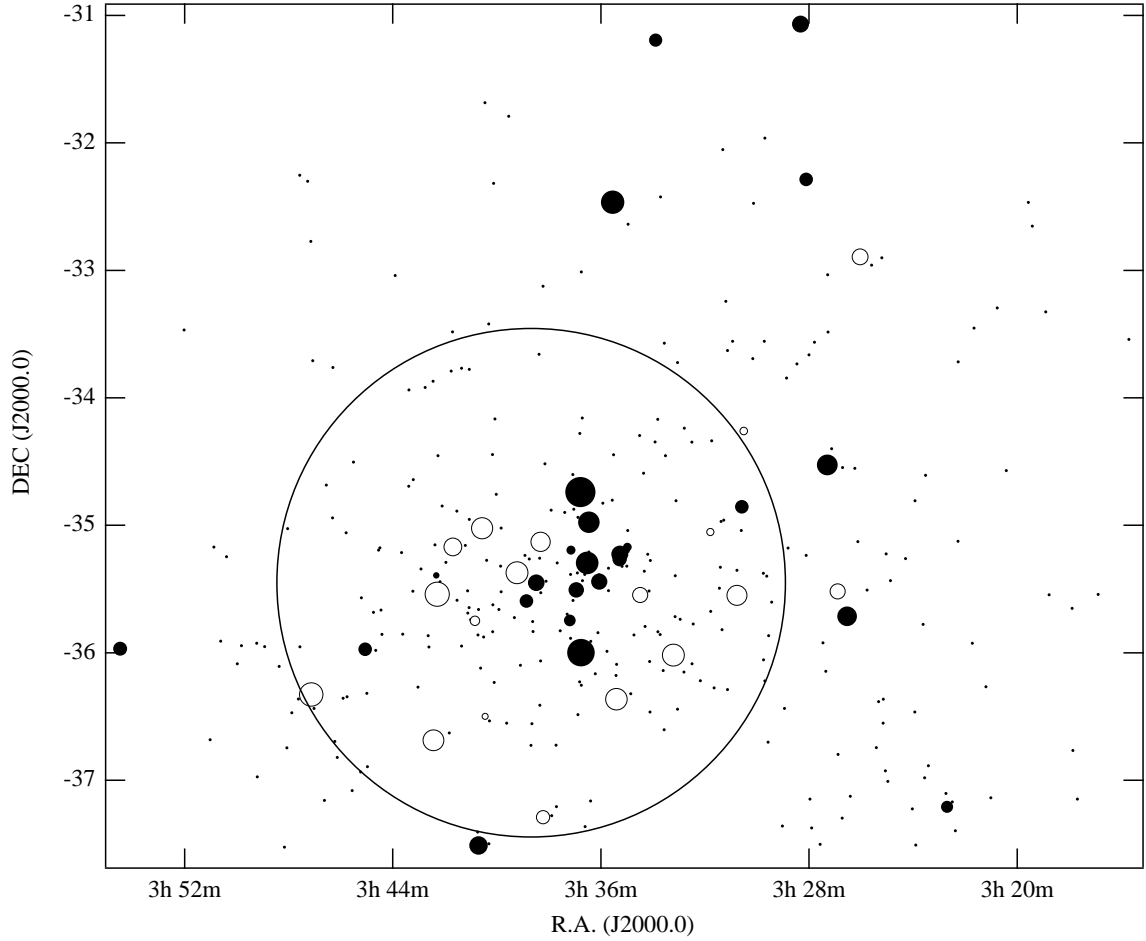}
\caption{The galaxy distribution in the Fornax cluster based on the Fornax Cluster Catalog \citep{F89}. All galaxies are shown as dots except the 43 early-type galaxies with available SBF distance measurements. The symbol size is inversely proportional to the galaxy distance. Dwarfs are open and giants are filled circles. The cluster core region is highlighted with a circle centered at NGC1399 and of radius two degrees equivalent to three cluster core radii or 720\,kpc.\label{distribution}}
\end{center}
\end{figure}

The Fornax cluster is less rich than Virgo and considered structurally to be less complex. It hosts about 350 likely cluster members \citep{F89}. They are essentially concentrated in one major component that is dominated by early-type galaxies (see Fig.~\ref{distribution}). The cluster has a King profile core radius of $R_{\rm c}=0.67^\circ$ ($\simeq 0.24$\,Mpc) and a center that almost exactly coincides with the location of the giant elliptical NGC1399 \citep{F89}. Approximately three degrees to the southwest lies a loose aggregate that is dominated by the peculiar Sa galaxy Fornax A (NGC1316) and composed of mainly late-type galaxies. In Ferguson's initial study, cluster membership and background objects were distinguished for the most part based on morphology. Subsequent spectroscopic surveys have reassigned a few members from likely membership to background status and vice versa and have also determined some previously unknown redshifts \citep{D98,SD01,D01}. \cite{M99} measured a relatively smooth Gaussian velocity distribution with very little substructure compared to Virgo. \cite{DG01} analysed redshifts of 108 Fornax cluster members, including 55 dwarfs, and found kinematical evidence for a large main concentration, centered around the velocity of NGC1399. The small second component around NGC1316 has a slightly higher ($\approx 100$\,km\,s$^{-1}$) mean velocity suggesting an infall towards the  cluster center from the near side. But all this substructure evidence is based on velocities only and thus does not provide a genuine three dimensional picture of the Fornax cluster.

Fornax was also subject to some initial distance work. The Cepheid distances of three spiral galaxies were measured, NGC1365 at 18.3$\pm2.3$\,Mpc, NGC1326A at 18.7$\pm1.9$\,Mpc \citep{S99}, and NGC1425 at 22.2$\pm1.4$\,Mpc \citep{M00}, only one of them (NGC1365) inside the central area as defined in Fig.~\ref{distribution}. These results provided some first insight into the spatial depth of the cluster. In order to get a better understanding of the galaxy distribution in the cluster core, \cite{J03} determined SBF distances to eight dEs and found a mean distance of $(m-M)_0=31.54\pm0.07$\,mag, or 20.3$\pm$0.7\,Mpc and a cluster depth of $\sigma_{\rm int}=1.4{+0.5 \atop -0.8}$\,Mpc. The former value was in good agreement with previous SBF work on 26 Fornax E/S0, Sa giants \citep[31.50$\pm0.04$\,mag,][]{T01} spread over a larger cluster area.

\begin{figure}
\begin{center}
\includegraphics[totalheight=0.7\textheight]{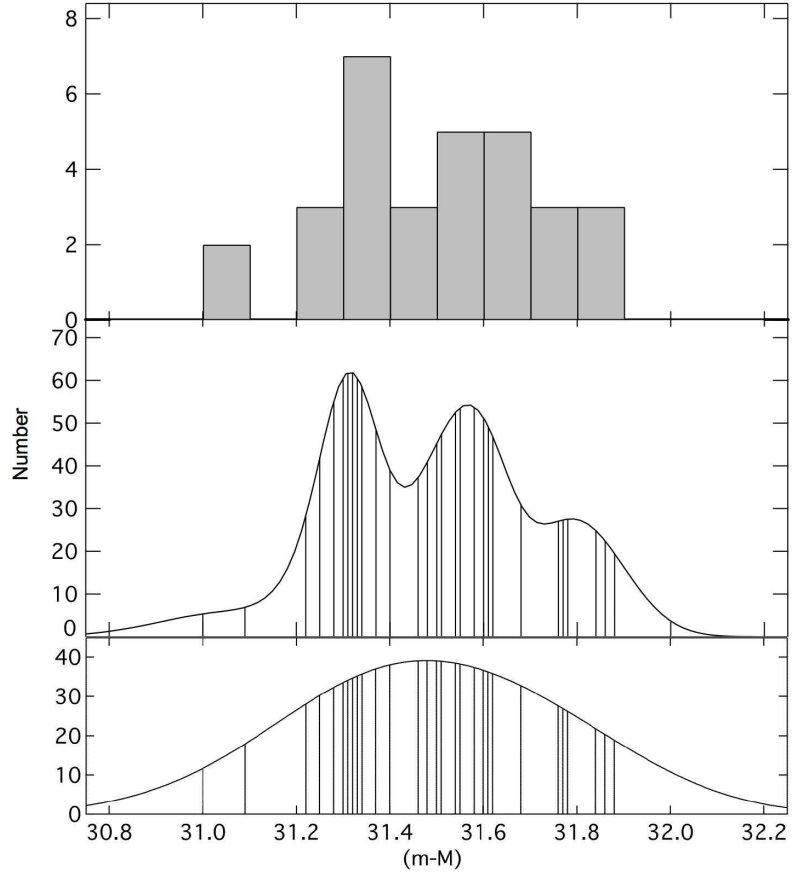}
\caption{The top figure shows a histogram of the distance distribution of the 31 dE+E galaxies within 2 degrees (720\,kpc) radius of NGC1399 with SBF distances from our studies and the literature. The second panel used an adaptive kernel method to test for substructure. The vertical lines apply to the individual galaxies used in this analysis. The third panel also used an adaptive kernel but adopts a 0.2\,mag average error and this appears to smooth the previously apparent substructure out.\label{ddist}}
\end{center}
\end{figure}

With the availability of SBF distances for a total of 43 early-type galaxies (including the 10 dEs from this study) we have the first-time luxury to conduct a 3-D Fornax cluster analysis right in the cluster center. Fig.~\ref{ddist} shows three different representations of the distance distribution of the 31 dE+E galaxies that are in projection closer than three core radii ($2^\circ$) from NGC1399. This sample size is equivalent to 35 percent of all early-type galaxies in the Fornax Cluster Catalog brighter than $-13.5$\,mag located in that area. The top panel is the histogram, binned in 0.1\,mag bins.  The second panel uses the adaptive kernel method \citep{V94} with a Gaussian kernel.  Interestingly, both graphs show evidence for two or even three concentrations, the two major ones centered at a distance modulus of $\approx$ 31.3 and at just under 31.6. The bottom panel is probably the closest to reality showing the distribution using the adaptive kernel method and adopting an average SBF distance uncertainty of 0.2\,mag.  The process completely straightens out the substructure indicating that the number of galaxies and the accuracy in the distance measurements are not sufficient to establish the reality of this effect. Nevertheless, the concentrations in the distribution of the individual galaxies still stand out and should be tested in future studies. The selected 31 dE+E galaxies give a preliminary distance of $19.95\pm0.37$\,Mpc [$(m-M)_0=31.49\pm0.04$\,mag] to the Fornax cluster core. A final value will be given in \S\ref{fdist}.

Next we wish to investigate the line-of-sight cluster depth to further reduce the uncertainties in the \citet{J03} result. 
The observed dispersion in the distance distribution of our sample galaxies, $\sigma_{(m-M)}=0.24$\,mag is influenced  by 
two factors: the distance errors and the intrinsic cluster depth. The depth can be calculated via: 
$\sigma_{\rm los}^2=\sigma_{(m-M)}^2-\sigma_{\Delta(m-M)}^2$, where $\sigma_{\Delta(m-M)}$ must be a representative 
estimate of the distance errors. Because the distance accuracy of our sample shows quite some variation we 
generated 100,000 bootstrap samples \citep{E86} of distances and errors from the original data set. For each sample we calculated 
its mean distance and mean error and used the previous equation to determine $\sigma_{\rm los}$. 
Histograms were then produced for all three parameters, each well approximated by a Gaussian. The $\sigma_{\rm los}^2$ distribution 
was found to peak at 0.0075 with a 1$\sigma$ scatter of 0.0105.  This translates into a true line-of-sight cluster depth 
of $\sigma_{\rm int}=0.74{+0.52 \atop -0.74}$\,Mpc  which is in good agreement with the small linear extension of Fornax in the sky
as estimated from the positions of early-type galaxies in the FCC: $\sigma_{\rm R.A.}=0.55\pm0.06$\,Mpc; 
$\sigma_{\rm DEC}=0.46\pm0.05$\,Mpc. We conclude that the early-type galaxy population in the central region of 
the Fornax cluster is spatially well constrained with no evidence of elongation along the line-of-sight. 

\subsection{The Hubble diagram} \label{hub}
\begin{figure}
\begin{center}
\includegraphics[totalheight=0.6\textheight]{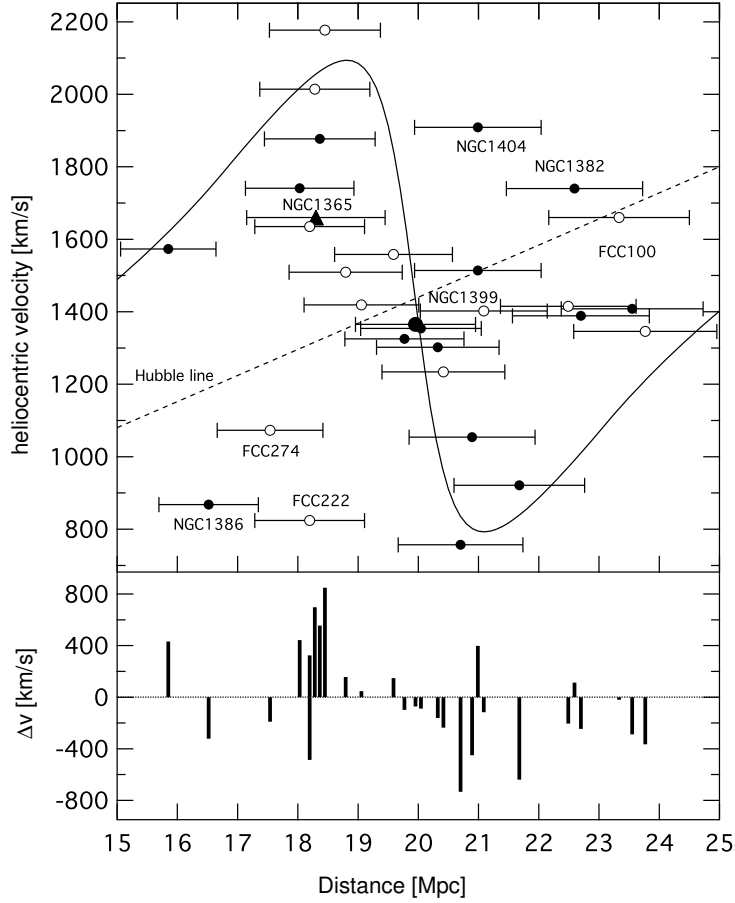}
\caption{{\em Top:} Velocity-distance diagram for 29 dE+E galaxies from the central Fornax cluster region. Open circles refer to dwarf ellipticals, filled circles represent giant ellipticals. Error bars correspond to 0.5$\sigma$ distance uncertainty. The Hubble line is indicated as a dashed line and the solid S-curve represents the best-fitting mass-age cluster model. Peculiar velocities are plotted relative to the Hubble expansion. For further discussion see text. The spiral galaxy NGC1365 that is located in the central cluster area and has a Cepheids distance is shown as a filled triangle. {\em Bottom:} Individual galaxy peculiar velocities are plotted as a function of distance.\label{hubbledia}}
\end{center}
\end{figure}

Unlike previous work \citep[e.g.][]{DG01} that concentrated on 
measuring global kinematical properties of the Fornax cluster 
we now have velocities and distances available for 29 early-type 
galaxies within a 2 degree (720\,kpc) radius from NGC1399, a data 
set highly suitable for analysing gravitationally induced motion right 
in the core of the Fornax cluster. The velocity-distance diagram for 
these galaxies is shown in Fig.\ref{hubbledia}.

On the near side, almost all galaxies (9 out of 12) with distance 
estimates closer than the tentative cluster center are redshifted 
relative to the cosmological expansion. Only the three galaxies NGC1386, FCC222, 
and FCC278 stand out as they appear isolated in the velocity-distance 
space. They are blueshifted by $\approx 360$\,km\,s$^{-1}$ relative to 
the Hubble flow and $\approx 700$\,km\,s$^{-1}$ to the the other sample 
galaxies at comparable distances. The main galaxy in the ensemble,
NGC1386, whose SBF distance was measured by \cite{T01}, is a 
Sa galaxy hosting a Seyfert 2 nucleus \citep{W91}. FCC222 was  
classified as a SAB(s)0 galaxy but no evidence for a disk or spiral 
features was found in our high resolution FORS1 images. Hence, the 
morphological type ``dE0,N" as initially given in the 
Fornax cluster catalogue \citep{F89} must be considered to be a more 
accurate description.

The situation in the Hubble diagram is equally pronounced on the 
far side of the cluster. With only two and possibly a third exception 
all galaxies are blueshifted to the systemic velocity. Those are 
NGC1404 together with NGC1382 and the dwarf FCC100. 
NGC1404 is a well known object in the X-ray community \citep{JS97,MD05}. 
The X-ray emission of NGC1404 shows a distorted envelope towards NGC1399, 
suggesting that NGC1404, which has a similar distance as NGC1399 but a very 
different velocity, experiences ram pressure stripping of the galaxy gas, caused by its 
relative motion of $\Delta v=544$\,km\,s$^{-1}$ through the surrounding 
intracluster medium. Furthermore, \cite{DG01} speculated that another galaxy, the irregular  
NGC1427A at $03^h 40^m09.3^s$ $-35^d37^m28^s$ (J2000) with a 
velocity $v_\odot=2028$\,km\,s$^{-1}$ may be associated to NGC1404 but
no distance is available to clarify this to date.

Overall the bulk of sample galaxies exhibit a prominent S-shaped infall pattern 
suggesting that Fornax is still in the process of formation during the present 
epoch through a general collapse. The Hubble line intercepts the kinematically
defined curve at a distance of $\approx 19.9$\,Mpc right at the location, 
i.e.~distance and velocity, of NGC1399 thereby highlighting the central 
role of this galaxy in the cluster.
Based on the prescription of \citet{T84} we used a point mass model to 
estimate the total cluster mass. Adopting 13.7\,Gyr  as the age of the 
Universe \citep[WMAP, ][]{S03} we fitted a mass model at the data points. 
The best-fitting envelope for the line of sight angle of two degrees from 
the Fornax cluster center is indicated as solid curve in Fig.~\ref{hubbledia} mapping 
the extreme model velocity. From the amplitude of $735$\,km\,s$^{-1}$ 
we get a cluster mass of $2.3\pm0.3\times 10^{14}\mathcal{M}_\odot$ within 
720\,kpc projected distance from NGC1399. This result is two times larger 
than the previously estimated Fornax mass by \citet{DG01} based 
solely on redshifts for early-type galaxies, and approximately 40 times the 
mass of NGC1399 \citep[$0.6\times 10^{13}\mathcal{M}_\odot$, ][]{JS97}. 
The estimated collapse time for the system is $t_{\rm collapse}=2.9{+1.6 \atop -0.9}$\,Gyr. 

\subsection{Fornax cluster distance and the Hubble constant} \label{fdist}
To determine the physically most meaningful distance to the Fornax cluster we need to remove the six outlying galaxies. The remaining selection of 23 early-type galaxies within a 2 degree radius of NGC1399 that follow the overall infall pattern of the cluster yields a true cluster distance of $20.13\pm0.40$\,Mpc [$(m-M)_0=31.51\pm0.04$\,mag]. This value shall now be used to estimate the Hubble constant. The standard procedure is to transform the mean velocity for the sample galaxies, $1447\pm 66$\,km\,s$^{-1}$, to the frame of the cosmic microwave background. This required two corrections: (i) a correction of $-115\pm 18$\,km\,s$^{-1}$ from heliocentric to the barycenter of the Local Group \citep{C99}, and (ii) a projected Virgocentric flow correction of $-54\pm 9$\,km\,s$^{-1}$ in the direction of Fornax. The latter is based on the distance geometry and a $1/R$ flow-field characteristics of the Local Group-Virgo-Fornax system as outlined in \citet{M99} and the assumption of an infall velocity of $240\pm40$\,km\,s$^{-1}$ of the Local Group into the Virgo cluster \citep{J93}. Finally we use the derived cosmological expansion velocity $v_{CMB}^{Fornax}=1278\pm69$\,km\,s$^{-1}$ to compute the Hubble constant of $63\pm5$km s$^{-1}$ Mpc$^{-1}$.  We find good agreement with the values derived from various independent techniques such as Cepheid stars by the HST ``Key Project'' team \citep[$72\pm8$\,km s$^{-1}$ Mpc$^{-1}$, ][]{F01}, the Cosmic Microwave Background by the WMAP team ($73\pm3$\,km s$^{-1}$ Mpc$^{-1}$), and galaxy clusters in the local Universe \citep[$69\pm5$km s$^{-1}$ Mpc$^{-1}$, ][]{G97}.

 \section{Summary} \label{conc}
In a few years time, HST will cease operation while the launch date for the next generation space telescope JWST is still uncertain. Without a facility to resolve galaxies into stars, the importance of the tip of the red giant branch method as extragalactic distance indicator will fade quickly. That situation will herald a new era for the surface brightness fluctuation method. Imaging surveys with wide-field cameras at optical and near-IR telescopes will cover large areas of sky that can be analysed with the SBF method to measure distances to thousands of early-type galaxies in a similar fashion as redshift surveys are carried out today. To prepare for that opportunity we are developing SAPAC, a fast and semi-automated SBF reduction software package. 

We employed SAPAC to process $B$ and $R$-band VLT+FORS1 images of 10 dE galaxies that were proposed members of the Fornax cluster.  The SBF distances confirm membership for all systems. We combined our results with SBF distances and velocities available for other Fornax dE+E galaxies to investigate the innermost region of the cluster. The derived Hubble diagram for 29 early-type galaxies closer than three cluster core radii ($2^\circ$) from the central galaxy NGC1399 exhibits a pronounced S-shaped infall pattern suggesting that Fornax is dynamically young in a phase of general collapse and possible accretion of a few individual galaxies. A cluster mass model gives the Fornax cluster mass of $2.3\pm0.3\times 10^{14}\mathcal{M}_\odot$ (within 720\,kpc projected distance from NGC1399) and the associated dynamical collapse time is $t_{\rm collapse}=2.9{+1.6 \atop -0.9}$\,Gyr. A kinematically cleaned galaxy subsample then yielded the true distance of the Fornax cluster core at 20.13$\pm$0.40\,Mpc [$(m-M)_0=31.51\pm 0.04$\,mag]. Applying a bootstrap resampling technique on the distance distribution further reveals a cluster depth of $\sigma_{\rm int}=0.74{+0.52 \atop -0.74}$\,Mpc consistent with 
the linear extension of the cluster in R.A.~and DEC. We conclude that the early-type galaxy population in the Fornax cluster is spatially well constrained
by a spherical shape with a characteristic radius of $\approx$0.6\,Mpc and we find marginal evidence for some spatial clumping. Combining the cosmological velocity of Fornax and the distance from the kinematically cleansed central cluster sample yields a Hubble constant of $H_0 = 63\pm 5$ \kms Mpc$^{-1}$. 

\begin{acknowledgements}
The authors thank the referee for carefully reading our manuscript and making helpful suggestions to improve our original paper.
They are also grateful to the anonymous ESO service observers at the VLT for obtaining the excellent quality images that were essential
for this study.
\end{acknowledgements}

%

\clearpage

\clearpage

\clearpage

\clearpage

\clearpage

\clearpage

\clearpage


\clearpage

\begin{deluxetable}{cccccccc}
\tablewidth{0pt}
\tabletypesize{\small}
\tablecaption{Basic properties of the observed early-type dwarf galaxies in the Fornax cluster. \label{basicdata}}
\tablecolumns{8}
\tablehead{\colhead{} & \colhead{} & \colhead{R.A.} &\colhead{Decl.} & \colhead{$B_T$} & \colhead{$v_\odot$} & \colhead{$A_B$} & \colhead{$A_R$} \\
\colhead{FCC} & \colhead{Type} & \colhead{(J2000.0)} & \colhead{(J2000.0)} & \colhead{(mag)} & \colhead{(\kms)} & \colhead{(mag)} & \colhead{(mag)} \\
\colhead{(1)} & \colhead{(2)} & \colhead{(3)} & \colhead{(4)}  & \colhead{(5)} & \colhead{(6)} & \colhead{(7)} & \colhead{(8)}}
\startdata
206 & dE0pec  & 03h38m13.4s & $-$37d17m18s & 15.8 & 1402$^\ast$ & 0.061 & 0.038\\
207 & dE2,N  & 03h38m16.2s & $-$37d07m39s  & 15.9 & 1419$^\ast$ & 0.063 & 0.039\\
222 & dE0,N  & 03h39m13.3s & $-$35d22m11s  & 15.6 &  953$^\ddagger$ & 0.056 & 0.035\\
243 & dE1,N  & 03h40m26.9s & $-$36d29m51s  & 16.5 & 1404$^\ast$ & 0.047 & 0.029\\ 
245 & dE0,N  & 03h40m33.7s & $-$35d01m17s  & 16.0 & 2124$^\ast$ & 0.051 & 0.032\\
252 & dE0,N  & 03h40m50.2s & $-$35d44m49s  & 16.0 & 1415$^\ast$ & 0.045 & 0.028\\
266 & dE0,N  & 03h41m41.2s & $-$35d10m08s  & 15.9 & 1551$^\ast$ & 0.045 & 0.028\\
274 & dE0,N  & 03h42m17.1s & $-$35d32m21s  & 16.5 & 1073$^\dagger$ & 0.050 & 0.031\\
279 & dE0,N  & 03h42m26.2s & $-$36d41m09s  & 16.7 & ---  & 0.032 & 0.02\\
318 & dE2,N  & 03h47m08.0s & $-$36d19m32s  & 16.1 & ---  & 0.051 & 0.032\\
\enddata
\tablerefs{ $\ast$ \citet{D01}, $\ddagger$ \citet{F89}, $\dagger$ \citet{H99}.  All data in columns 7 \& 8 from \citet{S98}. All data in columns 1, 2 and 5 from \citet{F89}}.
\end{deluxetable}

\clearpage

\begin{deluxetable}{ccccccccccc}
\tabletypesize{\scriptsize}
\tablecaption{Observing log of the imaging data taken in service mode.\label{logbook}}
\tablecolumns{11}
\tablehead{\colhead{}&\colhead{}&\colhead{}&\colhead{Mean}&\colhead{Exptime}&\colhead{FWHM}&\colhead{}&\colhead{}&\colhead{Mean}&\colhead{Exptime}&\colhead{FWHM}\\
\colhead{FCC}&\colhead{Filter}&\colhead{UT Date}&\colhead{Airmass}&\colhead{(sec)}&\colhead{(arcsec)}&\colhead{Filter}&\colhead{UT Date}&\colhead{Airmass}&\colhead{(sec)}&\colhead{(arcsec)}\\
\colhead{(1)}&\colhead{(2)}&\colhead{(3)}&\colhead{(4)}&\colhead{(5)}&\colhead{(6)}&\colhead{(7)}&\colhead{(8)}&\colhead{(9)}&\colhead{(10)}&\colhead{(11)}}
\startdata
206 & $B$ &  2001-11-20 & 1.03 & 3x500 & 0.69 & $R$ & 2002-08-13 & 1.29 & 3x700 & 0.55 \\ 
207 & $B$ &  2001-11-20 & 1.02 & 3x500 & 0.52 & $R$ & 2002-08-13 & 1.16 & 3x700 & 0.62 \\ 
222 & $B$ &  2001-12-07 & 1.02 & 3x500 & 0.91 & $R$ & 2002-08-13 & 1.09 & 3x700 & 0.58 \\ 
243 & $B$ &  2001-12-07 & 1.03 & 3x500 & 0.82 & $R$ & 2002-08-14 & 1.36 & 3x700 & 0.69 \\ 
245 & $B$ &  2001-12-07 & 1.05 & 3x500 & 0.79 & $R$ & 2003-08-07 & 1.03 & 6x700 & 0.69 \\ 
252 & $B$ &  2001-12-11 & 1.10 & 3x500 & 0.79 & $R$ & 2003-09-06 & 1.02 & 3x700 & 0.55 \\ 
266 & $B$ &  2001-12-11 & 1.18 & 3x500 & 0.98 & $R$ & 2003-08-06 & 1.12 & 3x700 & 0.70 \\ 
274 & $B$ &  2002-01-15 & 1.02 & 3x500 & 0.85 & $R$ & 2003-08-31 & 1.01 & 1x700 & 0.56 \\ 
279 & $B$ &  2002-01-12 & 1.02 & 3x500 & 0.94 & $R$ & 2003-09-29 & 1.34 & 3x700 & 0.79 \\ 
318 & $B$ &  2001-11-18 & 1.12 & 3x500 & 0.72 & $R$ & 2003-09-30 & 1.02 & 3x700 & 0.57\\
\enddata
\end{deluxetable}

\clearpage

\begin{deluxetable}{cccccc}
\tablewidth{0pt}
\tablecaption{Photometric parameters for each observing night\label{cterm}}
\tablecolumns{5}
\tablehead{\colhead{UT Date}&\colhead{$k_B$}&\colhead{$k_R$}&\colhead{$C_{B-R}$}&\colhead{$ZP$}\\
\colhead{(1)}&\colhead{(2)}&\colhead{(3)}&\colhead{(4)}&\colhead{(5)}}
\startdata
2001-11-18&0.27&0.12&$-$0.054&$27.069\pm(0.010)$(B)\\
2001-11-20&0.27&0.12&$-$0.054&$27.055\pm(0.005)$(B)\\
2001-12-07&0.29&0.12&$-$0.060&$27.038\pm(0.011)$(B)\\
2001-12-11&0.29&0.12&$-$0.060&$27.119\pm(0.005)$(B)\\
2002-01-12&0.30&0.13&$-$0.066&$27.058\pm(0.004)$(B)\\
2002-01-15&0.30&0.13&$-$0.066&$27.024\pm(0.005)$(B)\\
2002-08-13&0.21&0.07&$-$0.048&$27.210\pm(0.006)$(R)\\
2002-08-14&0.21&0.07&$-$0.048&$27.266\pm(0.033)$(R)\\
2003-08-06&0.24&0.11&$-$0.067&$27.347\pm(0.003)$(R)\\
2003-08-07&0.24&0.11&$-$0.067&$27.383\pm(0.002)$(R)\\
2003-09-01&0.24&0.11&$-$0.067&$27.163\pm(0.010)$(R)\\
2003-09-06&0.24&0.11&$-$0.067&$27.440\pm(0.016)$(R)\\
2003-09-29&0.24&0.11&$-$0.067&$27.328\pm(0.006)$(R)\\
2003-09-30&0.24&0.11&$-$0.067&$27.342\pm(0.009)$(R)\\
\enddata
\end{deluxetable}

\clearpage

\begin{deluxetable}{ccccccccc}
\tabletypesize{\scriptsize}
\tablewidth{0pt}
\tablecolumns{9}
\tablecaption{Parameters of SBF fields in two galaxies\label{fieldparam}}
\tablehead{\colhead{Galaxy} & \colhead{size} & \colhead{$m_1$} & \colhead{$\overline{g}$} & \colhead{$s$} & \colhead{$P_0$} & \colhead{$P_1$} & \colhead{$S/N$} & \colhead{$P_{BG}/P_0$} \\ \colhead{Field name} & \colhead{(pixels)} & \colhead{(mag)} & \colhead{(ADU)} & \colhead{(ADU)} & \colhead{(ADU s$^{-1}$ pixel$^{-1}$)} & \colhead{(ADU s$^{-1}$ pixel$^{-1}$)} &\colhead{ } &  \\ \colhead{(1)} & \colhead{(2)} & \colhead{(3)} & \colhead{(4)} & \colhead{(5)} & \colhead{(6)} & \colhead{(7)} & \colhead{(8)} &\colhead{ (9)}}
\startdata
FCC222(1,1)&70&27.11& 881.8& 9236.7&0.058(0.002)&0.008& 5.3&0.04\\
FCC222(1,2)&  &     &1370.2&       &0.084(0.002)&0.005&10.9&0.03\\
FCC222(1,3)&  &     & 718.0&       &0.087(0.004)&0.009& 7.4&0.03\\
FCC222(2,1)&  &     &1587.1&       &0.074(0.002)&0.004&10.2&0.03\\
FCC222(2,2)&  &     &4629.1&       &0.067(0.002)&0.002&14.4&0.04\\
FCC222(2,3)&  &     &1587.5&       &0.082(0.002)&0.004&11.6&0.03\\
FCC222(3,1)&  &     & 762.8&       &0.082(0.002)&0.009& 7.0&0.03\\
FCC222(3,2)&  &     &1505.6&       &0.088(0.003)&0.005&11.5&0.03\\
FCC222(3,3)&  &     & 943.7&       &0.070(0.004)&0.007& 7.0&0.04\\
FCC243(1,1)&54&27.19& 543.3&11655.3&0.080(0.004)&0.013& 5.0&0.04\\
FCC243(1,2)&  &     &1014.7&       &0.062(0.003)&0.008& 5.2&0.05\\
FCC243(1,3)&  &     & 489.6&       &0.063(0.004)&0.013& 3.8&0.05\\
FCC243(2,1)&  &     &1063.8&       &0.075(0.004)&0.007& 7.1&0.04\\
FCC243(2,2)&  &     &2754.7&       &0.072(0.005)&0.003&11.2&0.04\\
FCC243(2,3)&  &     & 815.9&       &0.114(0.006)&0.009& 9.3&0.03\\
FCC243(3,1)&  &     & 487.5&       &0.090(0.005)&0.014& 5.0&0.03\\
FCC243(3,2)&  &     & 778.2&       &0.087(0.006)&0.010& 6.5&0.03\\
\enddata
\end{deluxetable}

\clearpage

\begin{deluxetable}{cccc}
\tabletypesize{\scriptsize}
\tablewidth{0pt}
\tablecolumns{4}
\tablecaption{Fluctuation magnitudes and colours for SBF fields in two galaxies \label{fieldmag}}
\tablehead{\colhead{Galaxy} & \colhead{$A_R$} & \colhead{$\overline{m}_R^0$} &  \colhead{$(B-R)_0$} \\ \colhead{Field} & \colhead{(mag)} & \colhead{(mag)} & \colhead{(mag)} \\ \colhead{(1)} & \colhead{(2)} & \colhead{(3)} & \colhead{(4)}}
\startdata
%
%
FCC222(1,1)&0.03$\pm$0.01&30.20$\pm$0.04&1.25$\pm$0.03\\
FCC222(1,2)&             &29.78$\pm$0.03&1.22$\pm$0.03\\
FCC222(1,3)&             &29.75$\pm$0.05&1.27$\pm$0.04\\
FCC222(2,1)&             &29.93$\pm$0.03&1.23$\pm$0.03\\
FCC222(2,2)&             &30.04$\pm$0.04&1.23$\pm$0.03\\
FCC222(2,3)&             &29.81$\pm$0.04&1.23$\pm$0.03\\
FCC222(3,1)&             &29.81$\pm$0.04&1.26$\pm$0.04\\
FCC222(3,2)&             &29.73$\pm$0.04&1.22$\pm$0.03\\
FCC222(3,3)&             &29.99$\pm$0.06&1.25$\pm$0.03\\
FCC243(1,1)&0.03$\pm$0.01&29.93$\pm$0.07&1.14$\pm$0.04\\
FCC243(1,2)&             &30.22$\pm$0.06&1.15$\pm$0.03\\
FCC243(1,3)&             &30.20$\pm$0.09&1.15$\pm$0.04\\
FCC243(2,1)&             &30.00$\pm$0.06&1.15$\pm$0.03\\
FCC243(2,2)&             &30.05$\pm$0.07&1.18$\pm$0.03\\
FCC243(2,3)&             &29.54$\pm$0.06&1.15$\pm$0.03\\
FCC243(3,1)&             &29.81$\pm$0.07&1.15$\pm$0.04\\
FCC243(3,2)&             &29.85$\pm$0.08&1.14$\pm$0.03\\
\enddata
\end{deluxetable}

\clearpage

\begin{deluxetable}{ccc}
\tablewidth{0pt}
\tablecolumns{3}
\tablecaption{SBF distances for the Fornax dwarf ellipticals with two results listed for undecided cases (linear branch result listed first)\label{distlist}}
\tablehead{\colhead{FCC} & \colhead{$(m-M)_0$} & \colhead{D} \\ \colhead{Number} & \colhead{(mag)} & \colhead{(Mpc)} \\ \colhead{1} & \colhead{2} & \colhead{3}}
\startdata
206   &31.62$\pm$0.26&21.09$\pm$2.68\\
207   &31.40$\pm$0.13&19.05$\pm$1.18\\
\ldots&31.02$\pm$0.09&16.00$\pm$0.67\\
222   &31.30$\pm$0.18&18.20$\pm$1.57\\
243   &31.88$\pm$0.22&23.77$\pm$2.53\\
245   &31.33$\pm$0.18&18.45$\pm$1.59\\
252   &31.76$\pm$0.35&22.49$\pm$3.93\\
266   &31.46$\pm$0.33&19.59$\pm$3.21\\
274   &31.22$\pm$0.16&17.54$\pm$1.34\\
279   &31.34$\pm$0.14&18.54$\pm$1.23\\
318   &31.25$\pm$0.18&17.78$\pm$1.54\\
\ldots&31.06$\pm$0.10&16.29$\pm$0.77\\
\enddata
\end{deluxetable}

\end{document}